\documentclass[aps,prx,twocolumn,amsmath,amssymb,nofootinbib,superscriptaddress,floatfix,reprint,longbibliography]{revtex4-2}

\usepackage{tikz}
\usepackage{graphicx}
\usepackage{dcolumn}
\usepackage{bm}
\usepackage{braket}
\usepackage{xcolor}
\usepackage{hyperref}
\usepackage{booktabs}

\begin{document}



\title{Pareto Frontier of Neural Quantum States: Scalable, Affordable, and Accurate Convolutional Backflow for Strongly Correlated Lattice Fermions}


\author{Yuntian Gu}
\altaffiliation{The three authors contributed equally to this work.}
\affiliation{State Key Laboratory of General Artificial Intelligence, School of Intelligence Science and Technology, Peking University}
\affiliation{ByteDance Seed, China}

\author{Zeyao Han}
\altaffiliation{The three authors contributed equally to this work.}
\affiliation{Institute for Advanced Study, Tsinghua University, Beijing 100084, China}
\affiliation{ByteDance Seed, China}

\author{Wenrui Li}
\altaffiliation{The three authors contributed equally to this work.}
\affiliation{State Key Laboratory of General Artificial Intelligence, School of Intelligence Science and Technology, Peking University}
\affiliation{ByteDance Seed, China}

\author{Zhiyu Xiao}
\affiliation{Institute of Physics, Chinese Academy of Sciences}

\author{Tao Xiang}%
\email{txiang@iphy.ac.cn}
\affiliation{Institute of Physics, Chinese Academy of Sciences}

\author{Mingpu Qin}%
\email{qinmingpu@sjtu.edu.cn}
\affiliation{Key Laboratory of Artificial Structures and Quantum Control (Ministry of Education), School of Physics and Astronomy, Shanghai Jiao Tong University}

\author{Liwei Wang}%
\email{wanglw@pku.edu.cn}
\affiliation{State Key Laboratory of General Artificial Intelligence, School of Intelligence Science and Technology, Peking University}
 
\author{Dingshun Lv}%
\email{ywlds163@gmail.com}
\affiliation{ByteDance Seed, China}
\affiliation{FieldQuantum Research Institute, Beijing, 100084, China}




\date{\today}

\begin{abstract}
The recent development of Neural Quantum States (NQS) has established them as one of the most accurate methods for studying strongly correlated many-fermion systems, outperforming existing many-body approaches for large systems. However, NQS calculations are currently extremely resource-intensive. In this work, we introduce a new Pareto frontier of efficiency and accuracy for NQS in the simulation of strongly correlated lattice fermions. This frontier is defined by two complementary backflow-related architectures: the Sparse Convolutional Ansatz for Lattice Electrons (SCALE), which achieves state-of-the-art efficiency, and the Accurate Convolutional ansatz for lattice Electrons (ACE), which delivers state-of-the-art accuracy, based on benchmark results on the iconic Hubbard and $t-J$ models on large lattice sizes. SCALE utilizes a tailored convolutional design to enable a highly efficient local update based on the low-rank update of determinants. This structure reduces the computational scaling from $\mathcal{O}(N^4)$ to $\mathcal{O}(N^3)$ in backflow methods, resulting in a practical speed-up exceeding $40\times$ in our test cases while retaining exceptional variational accuracy. As an application, we study the $1/8$-doped pure Hubbard model up to size $32 \times 32$, which was previously inaccessible, and find no significant energy difference between the horizontal and vertical filled stripe states, in contrast to the half-filled stripe state when next-nearest-neighbor hoppings are included. ACE, conversely, employs a deep convolutional stack to maximize expressive power, achieving unprecedented accuracy on large systems. We perform extensive benchmark calculations on the Hubbard and $t-J$ models with the two new architectures. On the challenging repulsive Hubbard model, SCALE achieves variational energies competitive with leading methods, but at a fraction of the computational cost. Meanwhile, ACE establishes a new state-of-the-art accuracy, decisively surpassing recent benchmarks while requiring only one-sixth of the runtime for system size $16 \times 4$. 
These new NQS approaches provide \emph{scalable}, \emph{affordable}, and \emph{accurate} tools for exploring the rich physics of strongly correlated fermionic systems—such as the microscopic mechanism of unconventional superconductivity.
\end{abstract}

\maketitle


\section{Introduction}

Understanding the emergence of exotic physics in strongly correlated fermion systems is a central and persistent challenge in condensed matter physics. Except for rare cases \cite{PhysRevLett.20.1445}, exact analytical solutions are unattainable, necessitating the application of computational methods. However, strongly correlated systems usually harbor competing states with nearly degenerate low energies, a feature that poses significant challenges for numerical simulations \cite{RevModPhys.87.457,qin2022hubbard, arovas2022hubbard}. Consequently, results can be highly sensitive to finite-size effects and boundary conditions \cite{PhysRevB.102.041106,doi:10.1126/science.adh7691,shen2025ground,PhysRevLett.132.066002,jiang2025competitionchargedensitywavesuperconductingorders} as well as algorithmic biases, which may artificially favor one state over another. Over the past decades, a diverse array of numerical many-body methods has been developed, and benchmark calculations have been performed on iconic systems \cite{leblanc2015solutions}.
These methods collectively populate a Pareto front of computational tools, and the primary challenge remains to advance this frontier towards higher accuracy and greater efficiency.

Among existing numerical many-body methods, variational approaches are particularly powerful for determining the ground-state properties of strongly correlated quantum many-body systems. For instance, the density matrix renormalization group (DMRG) \cite{PhysRevLett.69.2863} method can provide nearly exact results for quasi-one-dimensional systems. However, the application of DMRG to two-dimensional systems suffers from the limited entanglement entropy encoded in the underlying wave-function ansatz, i.e., the Matrix Product States (MPS) \cite{PhysRevLett.75.3537,SCHOLLWOCK201196,Cirac2021rmp}. Although two-dimensional extension of MPS like projected-entangled-pair-states (PEPS) \cite{verstraete2004} can encode the entanglement area law of the ground state of two-dimensional systems, they are often hindered by high computational complexity \cite{orus_practical_2014,Cirac2021rmp,xiang2023density}. More recently, neural quantum states (NQS) have emerged as a highly flexible and expressive class of variational ansatz \cite{carleo2017solving}. 

To effectively incorporate neural networks, approaches that directly generalize mean-field theories, such as Neural Network Backflow (NNBF) \cite{luo2019backflow} and Hidden Fermion states \cite{robledo2022fermionic}, have proven to be highly accurate and efficient. Building upon these foundations, various architectures have been developed to better capture complex many-body correlations. For example, Transformer-Backflow utilizes an attention mechanism to effectively capture long-range correlations, surpassing the best tensor network methods for large systems \cite{gu2025solving}. Conversely, the Tensor-Backflow method \cite{liang2025investigating} generates the backflow Slater determinant based on tensor representation. Other specialized architectures, such as the Hidden Fermion Pfaffian State (HFPS) \cite{chen2025neural, roth2025superconductivitytwodimensionalhubbardmodel}, offer fresh perspectives on complex phases by suggesting that superconductivity is enhanced by the next-nearest-neighbor hopping $t^{\prime}$. Furthermore, recent studies have demonstrated that even simpler Multi-Layer Perceptron (MLP) network backbones can achieve significantly enhanced accuracy when augmented with rigorous symmetry projections \cite{loehr2025enhancingneuralnetworkbackflow}.

Despite these advancements, accurate NQS calculations remain extremely resource-intensive. Developing more efficient and accurate NQS architectures is crucial for enabling their widespread application in the study of strongly correlated many-electron systems. In this work, we introduce two complementary backflow-related architectures that define a new Pareto frontier for NQS simulations: the Sparse Convolutional Ansatz for Lattice Electrons (SCALE) and the Accurate Convolutional ansatz for lattice Electrons (ACE). 

SCALE is engineered for unprecedented efficiency while remaining one of the most accurate methods available. It employs a convolutional layer that processes only local regions of the input at each time. This locality is the key to a highly optimized local update scheme, which leverages cached intermediate computations and fast low-rank matrix updates. This design reduces the computational scaling of the simulation from $\mathcal{O}(N^4)$ to $\mathcal{O}(N^3)$, unlocking a practical speed-up that exceeds 40$\times$ for large systems. 
Complementing the efficient SCALE architecture, we introduce ACE to establish the pinnacle of accuracy on this frontier. ACE replaces SCALE's efficient layer with a deep stack of convolutional blocks, maximizing expressive power. While this design forfeits the fast update capability, it is built to further increase the accuracy over SCALE.

We perform extensive benchmark calculations on several canonical strongly correlated models. For the attractive Hubbard model, our results agree well with the intrinsically unbiased Determinant Quantum Monte Carlo (DQMC) simulation \cite{blankenbecler1981monte,hirsch1985two,PhysRevB.28.4059,PhysRevLett.123.136402}, and correctly identify the expected long-range s-wave pairing. On the more challenging repulsive Hubbard model, our framework demonstrates its full power. The efficient SCALE ansatz achieves a lower variational energy than several leading methods, including PEPS \cite{liu2025}, Tensor-Backflow \cite{liang2025investigating}, and HFPS~\cite{chen2025neural}. It is also directly competitive with the top-tier Transformer-Backflow~\cite{gu2025solving}, but at a fraction of the computational cost. Meanwhile, the ACE results establish a new accuracy record while remaining relatively efficient, decisively surpassing the best results of recent times \cite{loehr2025enhancingneuralnetworkbackflow}, while requiring only one-sixth of the computational cost. Crucially, the breakthrough efficiency of SCALE allows us to address an open question concerning the orientation of stripe state that was computationally prohibitive for previous methods \cite{gu2025solving}; our simulations on unprecedented $16 \times 32$ systems show that, in contrast to the partially filled stripe in the Hubbard model with next-nearest-neighboring hoppings \cite{gu2025solving}, we find no significant difference in the energies for the horizontal and vertical filled stripe in the pure Hubbard model with only nearest-neighboring hoppings. We further apply our method to the $t-J$ model, obtaining energies comparable to large-scale DMRG on narrow cylinders \cite{shen2025ground} and lower than the DMRG results for wide cylinders. These comprehensive results establish SCALE and ACE as robust and powerful new methods, defining the new state-of-the-art for both scalable and high-accuracy simulations of strongly correlated systems.

\section{Methods}

\subsection{Variational Monte Carlo}

The Variational Monte Carlo (VMC) method approximates the ground state of a quantum system by optimizing a parameterized trial wavefunction, or ansatz, $|\psi_{\theta}\rangle$. According to the variational principle, the energy expectation value $E(\theta) = \langle\psi_{\theta}|\hat{H}|\psi_{\theta}\rangle / \langle\psi_{\theta}|\psi_{\theta}\rangle$ provides a rigorous upper bound to the true ground state energy. In practice, this expectation value is computed stochastically by sampling configurations in the Fock basis $\{\bm n\}$ with respect to the probability distribution $P(\mathbf{n}) \propto |\psi_{\theta}(\mathbf{n})|^2$, where $\psi_\theta(\bm n) \equiv \braket{\bm n | \psi_\theta}$. This approach reformulates the energy expectation value as an average of the local energy, defined as $E_{\mathrm{loc}} = \sum_{\bm n'} \psi_\theta(\bm n)^{-1} \braket{\bm n | \hat{H} | \bm n'} \psi_\theta(\bm n')$, and can be obtained from configurations sampled using Markov chain Monte Carlo algorithms. The wavefunction parameters $\theta$ are then optimized using gradient-based methods to minimize the estimated energy, yielding the best variational approximation of the ground state. The details of the optimization method will be discussed in the following sections.


\begin{figure*}
\includegraphics[width=0.95\textwidth]{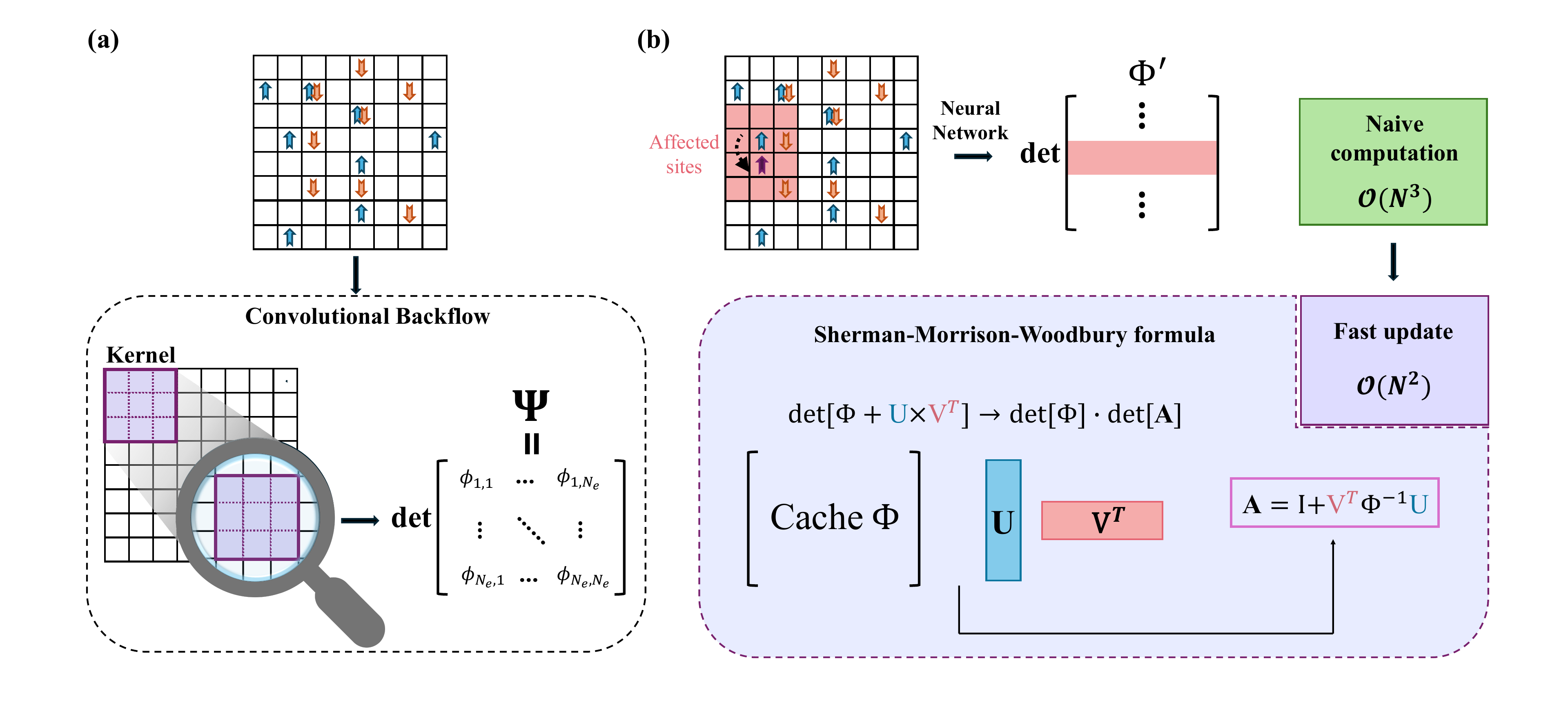}
\caption{\textbf{Network Architecture and Efficient Local Updates.} (a) Given an input configuration, a convolutional layer gathers information from local neighborhoods. The wavefunction is expressed as the determinant of the resulting backflow matrix. (b) SCALE enables efficient updates by only recomputing the orbitals of the affected sites. The computational cost of the determinant is reduced from $O(N^3)$ to $O(N^2)$ via the Sherman-Morrison-Woodbury formula.}
\label{fig:1}
\end{figure*}

\subsection{Sparse Convolutional Ansatz for Lattice Electrons (SCALE)}
\label{sec:SCALE}

In this section, we introduce a new NQS architecture that combines high expressive power with locality. A key requirement for efficient VMC simulations is the rapid calculation of wavefunction amplitude ratios, $\psi_{\theta}(\mathbf{n'}) / \psi_{\theta}(\mathbf{n})$, for configurations $\mathbf{n}$ and $\mathbf{n'}$ that differ only locally. This operation is fundamental not only to the MCMC sampling process but also to the evaluation of the local energy for Hamiltonians containing local operators, such as hopping terms in the Hubbard model. Our architecture, most directly inspired by FiRE \cite{scherbela2025accurate}, is designed explicitly to take advantage of this locality.

The wave-function ansatz takes the form of $\psi_{\theta}(\mathbf{n}) = \det[\Phi(\mathbf{n})]$, where the Slater matrix $\Phi$ is constructed from quasi-particle orbitals. These orbitals are not static but dynamically generated via a deep neural network that implements a backflow transformation \cite{PhysRev.46.1002,PhysRev.102.1189,PhysRevB.78.041101}, 
allowing the orbitals to depend on the particle configuration, and thereby capture complex electronic correlations. The generation of these configuration-dependent orbitals proceeds through a sequence of computational steps, as shown in Figure \ref{fig:1}(a).

\paragraph{Embedding Layer.}
Each lattice site $i \in \{1, ..., L_x \times L_y\}$ has a discrete physical state determined by its occupancy with spin degrees of freedom. For the Hubbard model, the local Hilbert space is 4-dimensional, corresponding to the states $\ket{0}$, $\ket{\uparrow}$, $\ket{\downarrow}$, and $\ket{\uparrow\downarrow}$. We first map these physical states to high-dimensional vectors using a learnable embedding matrix $\mathbf{E}_i \in \mathbb{R}^{4 \times h}$ where $h$ is the hidden dimension. Therefore, for a configuration $\mathbf{n}$, this gives an embedding matrix $\mathbf{H}^{(0)} = [\mathbf{E}_1(\mathbf{n}_1), ..., \mathbf{E}_{L_x\times L_y}(\mathbf{n}_{L_x \times L_y})]^\top \in \mathbb{R}^{(L_x \times L_y) \times h}$.

\paragraph{Convolutional Layer.}
The initial features are refined via a convolutional operation with residual connection~\cite{he2016deep}:
\begin{equation}
\label{eq:conv}
\mathbf{H}^{(1)} = \mathbf{H}^{(0)} + \sigma\left( \text{Conv}_{\theta_c}(\mathbf{H}^{(0)}) \right).
\end{equation}

Here, $\text{Conv}_{\theta_c}$ denotes a convolutional layer parameterized by $\theta_c = \{ \mathbf{K}, b \}$ (convolution kernel $\mathbf{K} \in \mathbb{R}^{k \times k \times h \times h}$ with kernel size $k$, bias $b \in \mathbb{R}^h$), operating with the same boundary conditions as the Hamiltonian. The $k\times k$ kernel specifies the local neighborhood, and $\sigma$ denotes the Sigmoid Linear Unit (SiLU) activation function \cite{elfwing2018sigmoid}. 

\paragraph{Site-wise MLPs.} Each site's feature vector undergoes transformation through a multi-layer perceptron (MLP) consisting of $L_{\text{MLP}}$ fully-connected layers, introducing non-linear expressivity through a sequence of affine transformations and activations. For layer indices $l = 1, ..., L_{\text{MLP}}$, the transformation follows:

\begin{equation}
\label{eq:MLP}
\mathbf{H}^{(l+1)} = \sigma\left( \mathbf{H}^{(l)}  \mathbf{W}^{(l)} + \mathbf{b}^{(l)} \right),
\end{equation}
where $\mathbf{W}^{(l)} \in \mathbb{R}^{h \times h}$ and $\mathbf{b}^{(l)} \in \mathbb{R}^h$ represent the weight matrix and the bias vector for the $l$-th layer, respectively.

\paragraph{Orbital Projection.} The final output $\mathbf{H}^{(L_{\text{MLP}} + 1)}$ is transformed by a linear layer to produce backflow orbitals $M \in \mathbb{R}^{2N \times N_e}$ where $N=L_x \times L_y$, $2$ denotes spin up and spin down and $N_e$ is the number of electrons. The Slater matrix $\Phi$ is generated by selecting the rows of $M$ corresponding to the occupied sites.

This architectural sequence ensures that orbitals are dynamically generated according to the particle configuration, capturing non-local correlations while maintaining the locality required for efficient wavefunction update.

\subsection{Efficient Local Updates}

The design of our ansatz enables a highly efficient local update scheme, leveraging both cached intermediate results and low-rank matrix update techniques as shown in Figure~\ref{fig:1}(b). In this section, we show the details of the mathematical formalism for these highly efficient implementations, which reduce the computational overhead of wavefunction updates during local energy evaluation and MCMC sampling.

\paragraph{Cached Computation for Neural Network Propagations.}

To avoid redundant calculations when the particle configuration undergoes local changes, we maintain a cache of intermediate network outputs. For a configuration update from $\mathbf{n}$ to $\mathbf{n}'$, the cache allows us to recompute only the affected components of the orbital generation pipeline. We first identify sites where occupancy changes:
   \begin{equation}
   \text{exchange} = \{ i \mid \mathbf{n}'_i \neq \mathbf{n}_i \}.
   \end{equation}

Due to the locality of convolutional layers, changes propagate only within a finite neighborhood determined by the size of the convolution kernel. For a convolutional kernel of size $k \times k$, the set of sites affected by changes in $\text{exchange}$ is:
   \begin{equation}
   \label{eq:neighbor}
   \text{affect\ sites} = \bigcup_{i \in \text{exchange}} \mathcal{N}_k(i),
   \end{equation}
   where $\mathcal{N}_k(i)$ denotes the neighborhood of site $i$ within Chebyshev distance $(k-1)/2$. Instead of re-evaluating the entire network, we only update features for $\text{affect\ sites}$. This caching strategy reduces the computational complexity of network updates from $\mathcal{O}(N)$ to $\mathcal{O}(|\text{affect site}|)$.

\paragraph{Fast Determinant Updates via Sherman–Morrison–Woodbury formula.}

For Slater matrix $\Phi(\mathbf{n}) \in \mathbb{R}^{N_e \times N_e}$, a local configuration change induces a low-rank modification \cite{PhysRevB.40.506}. We exploit this structure to update the determinants along with their inverses efficiently.

Let $\Phi' = \Phi(\mathbf{n}')$ denote the updated Slater matrix. The difference between  $\Phi'$ and $\Phi (= \Phi(\mathbf{n}))$ lies in the rows corresponding to the changed indices:
\begin{equation}
\Phi' = \Phi + \mathbf{U} \mathbf{V}^\top,
\end{equation}
where $\mathbf{U} \in \mathbb{R}^{N_e \times r}$ and $\mathbf{V} \in \mathbb{R}^{N_e \times r}$ with $r \ll N_e$. Using the Sherman–Morrison–Woodbury formula, the ratio of determinants and the inverse of the updated matrix are:
\begin{equation}
\begin{aligned}
\frac{\det(\Phi')}{\det(\Phi)} &= \det\left( \mathbf{A} \right)\\
(\Phi')^{-1} &= \Phi^{-1} - \Phi^{-1}\mathbf{U} \mathbf{A}^{-1} \mathbf{V}^\top \Phi^{-1},
\end{aligned}
\end{equation}
where $\mathbf{A} = \mathbf{I}_r + \mathbf{V}^\top \Phi^{-1} \mathbf{U}$. 
Critically, both the determinant and the inverse of $\mathbf{A}$ can be derived from a single LU decomposition, further optimizing the calculation.

This combined strategy of cached network computations and low-rank matrix updates reduces the per-update complexity significantly, enabling large-scale VMC simulations.

\subsection{Accurate Convolutional ansatz for lattice Electrons (ACE)}

We also introduce a second, computationally intensive new ansatz: the Accurate Convolutional ansatz for lattice Electrons (ACE), which can achieve state-of-the-art accuracy and establish definitive benchmarks. This architecture prioritizes maximal expressive power through increased network depth and a large neighborhood size, albeit at the cost of the efficient local update capabilities described in the previous section.

While ACE retains the overall structure of SCALE, it replaces the single convolutional layer (Eq. \ref{eq:conv}) with a deep stack of $L_{\text{CNN}}$ convolutional blocks. Starting from the initial embedding $\mathbf{H}^{(0)}$, the propagation through this deep stack proceeds as follows for each layer $l = 0, \dots, L_{\text{CNN}}-1$:
\begin{equation}
\mathbf{H}^{(l+1)} = \text{LN}\left(\mathbf{H}^{(l)} + \sigma\left( \text{Conv}_{\theta_c}^{(l)}(\mathbf{H}^{(l)}) \right) \right),
\end{equation}
where $\text{Conv}_{\theta_c}^{(l)}$ denotes the $l$-th convolutional layer and $\sigma$ is the SiLU activation. We employ Layer Normalization ($\text{LN}$)~\cite{ba2016layer} to stabilize the training across this deeper network. The final output $\mathbf{H}^{(L_{\text{CNN}})}$ is subsequently passed to the site-wise MLPs and orbital projection defined in Sec.~\ref{sec:SCALE}.

The primary trade-off for the increased accuracy in ACE is a reduction in computational efficiency. As the number of layers $L_{\text{CNN}}$ increases, the network's effective range expands significantly. In a sufficiently deep network, a local change at a single site $i$ propagates through the layers, altering the feature vectors $\mathbf{H}^{(L_{\text{CNN}})}$ across all lattice sites. This global dependency destroys the locality property required for the cached computation scheme (Eq. \ref{eq:neighbor}). Consequently, any update from configuration $\mathbf{n}$ to $\mathbf{n}'$ necessitates a full re-evaluation of the network pipeline. Thus, ACE serves as a powerful, high-accuracy architecture that can take advantage of the trade-off between expressivity and efficiency.

\subsection{Comparison to Existing Architectures}

To contextualize the advancements of SCALE and ACE, it is instructive to compare their fundamental design principles with other recent competitive approaches, such as Transformer-Backflow \cite{gu2025solving} and Hidden Fermion (HF) methods like HFPS \cite{chen2025neural}. The Hidden Fermion approach is conceptually completely distinct from the NNBF paradigm that underpins SCALE and ACE. Specifically, HF methods construct wavefunctions by evaluating Slater determinants in an augmented Hilbert space involving ``hidden'' additional fermionic degrees of freedom. These states are then projected back onto the physical Hilbert space through a constraint which is optimized using a neural network parametrization. In contrast, backflow-based methods operate strictly within the original physical Hilbert space, using neural networks to dynamically generate configuration-dependent, multi-particle orbitals.

Within the backflow paradigm, the defining novelty of SCALE is its resolution of the asymptotic scaling bottlenecks inherent to NQS. In standard NNBF architectures, the neural network introduces global dependencies: a local change in a single electron's position instantly alters the hidden features of all spatial orbitals. This inflicts a severe computational penalty, necessitating a full $\mathcal{O}(N)$ re-evaluation of the neural network for every single local update. By explicitly restricting the network's receptive field using a tailored convolutional geometry, SCALE ensures that physical updates only perturb a mathematically bounded local neighborhood. This architectural sparsity allows for exact caching, slashing the network forward-pass cost from $\mathcal{O}(N)$ down to an $\mathcal{O}(1)$ operation per update.

On the other hand, ACE offers a conceptual novelty that challenges current machine learning trends. While global attention mechanisms (such as Vision Transformers \cite{dosovitskiy2020image}) have become the default for 2D structured data, they effectively discard the strict spatial geometry of the lattice in favor of fully connected, all-to-all graphs. Reverting to deep convolutional stacks instead of attention is a highly non-trivial choice in modern ML, but it is uniquely suited to the problem domain. Strongly correlated lattice fermions are governed by strictly local operators (e.g., nearest-neighbor hopping $t$ and on-site repulsion $U$), meaning long-range correlations build up spatially. ACE respects this physical inductive bias by employing a deep stack of localized convolutional blocks. This forces the network to extract features hierarchically, matching the physical reality of the lattice, rather than relying on naive all-to-all attention. ACE demonstrates the fundamental physical insight that deep, geometrically constrained convolutions are actually more effective at propagating information and capturing ground-state correlations for these specific local Hamiltonians than global attention mechanisms. By doing so, ACE establishes a new pinnacle of variational accuracy, decisively outperforming both Transformer-Backflow and HFPS, while maintaining a lighter computational footprint than full-attention models.

\subsection{Optimization and Ground State Projection}

We use the pretrain method \cite{gu2025solving} to provide NQS with better initialization. The variational parameters of the wavefunction are then optimized within the VMC framework using the MARCH optimizer, a sophisticated stochastic gradient method designed for the stable and efficient training of NQS \cite{gu2025solving}. To further improve upon the variational result, we subsequently employ Green's Function Monte Carlo (GFMC), a projector method that filters out residual excited-state contributions from the optimized VMC wavefunction under the fixed-node approximation \cite{PhysRevB.41.4552, PhysRevB.51.13039}. Throughout this paper, we have ensured sufficiently large Monte Carlo sampling such that, unless explicitly denoted with parentheses, the statistical error bars are one or several orders of magnitude smaller than the precision of the reported numbers in all tables and smaller than the data markers in all figures.

\section{Results}

To illustrate the performance of SCALE and ACE, we perform extensive calculations for the two-dimensional Hubbard model \cite{J.Huardbard} and its low-energy effective Hamiltonian in the large repulsion interaction limit, the $t-J$ model. The Hamiltonians of the Hubbard model and the $t-J$ models are
\begin{equation}
\begin{aligned}
H_h = &-t\sum_{\langle i,j \rangle, \sigma} \left( c_{i\sigma}^\dagger c_{j\sigma} + \text{h.c.} \right)
- t'\sum _{\langle\langle i,j \rangle\rangle, \sigma} \left( c_{i\sigma}^\dagger c_{j\sigma} + \text{h.c.} \right) \\
&+ U \sum _i n_{i\uparrow} n_{i\downarrow}
\end{aligned}
\end{equation}
and
\begin{equation}
\begin{aligned}
H_{t-J} = &-t\sum_{\langle i,j \rangle, \sigma} \mathcal{P}_G \left( c_{i\sigma}^\dagger c_{j\sigma} + \text{h.c.} \right) \mathcal{P}_G\\
&- t'\sum _{\langle\langle i,j \rangle\rangle, \sigma} \mathcal{P}_G \left( c_{i\sigma}^\dagger c_{j\sigma} + \text{h.c.} \right) \mathcal{P}_G \\
&+ J\sum_{\langle i,j \rangle} \mathbf{S}_i \cdot \mathbf{S}_j
\end{aligned}
\label{ham-tj}
\end{equation}
respectively. In these Hamiltonians, $\langle ij \rangle$ denotes nearest-neighbor sites and $\langle\langle ij \rangle\rangle$ denotes next-nearest-neighbor sites. $c_{i\sigma} (c_{i\sigma}^\dagger)$ is the annihilation and creation operator for spin spices $\sigma$ at site $i$. $\mathbf{S}_i$ is the spin operator at site $i$. We set the nearest neighbor hopping amplitude $t=1$ as the unit of energy, and primarily focus on the $1/8$ hole doping. In the $t-J$ Hamiltonian, $\mathcal{P}_G = \prod_i (1 - n_{i\uparrow} n_{i\downarrow})$ is the Gutzwiller projection operator to remove double occupancy.

\subsection{Computational Complexity}

\begin{figure*}
\centering
\includegraphics[width=0.8\textwidth]{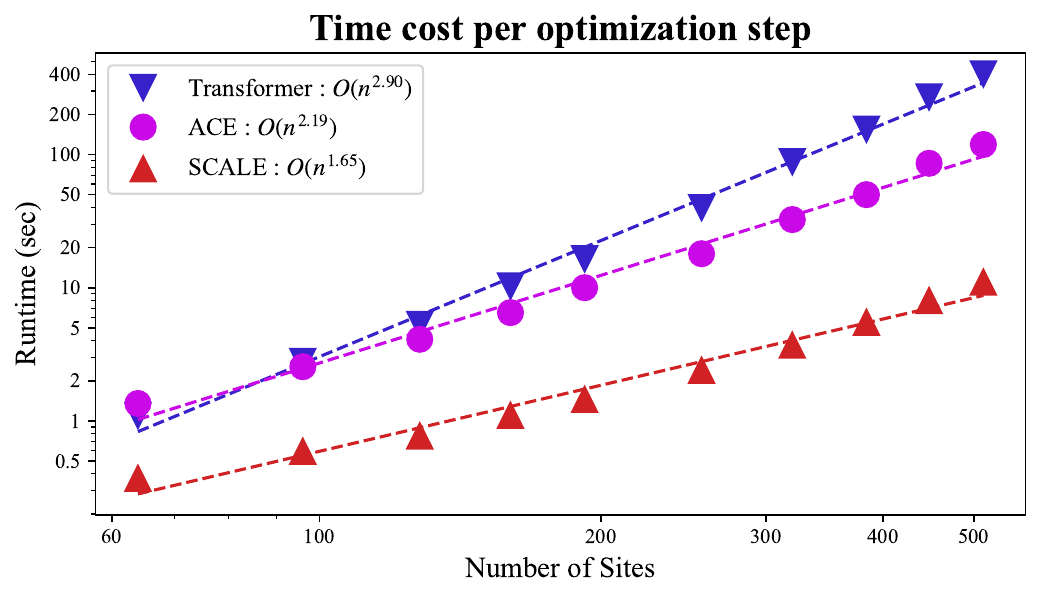}
\caption{\textbf{Time cost per optimization step as a function of system size.} The plot compares the runtime scaling of SCALE against global architectures (ACE and Transformer) under a fixed batch size. SCALE achieves a highly favorable asymptotic complexity (empirically $\approx \mathcal{O}(N^{1.65})$), resulting in substantial speed-ups for larger systems compared to the steeper scaling of ACE ($\approx \mathcal{O}(N^{2.19})$) and Transformer ($\approx \mathcal{O}(N^{2.90})$).}
\label{fig:2}
\end{figure*}

We first analyze the computational scaling of NQS by identifying the primary bottleneck: MCMC sampling and local energy evaluation, which requires $\mathcal{O}(N)$ steps of local update. In global architectures like ACE or Transformers, any local change to the configuration propagates through the entire network. This necessitates a full re-evaluation of the wavefunction for every proposed update. Since the determinant calculation scales as $\mathcal{O}(N^3)$, performing $\mathcal{O}(N)$ updates yields an asymptotic theoretical complexity of $\mathcal{O}(N^4)$ per sweep. However, in practice, the total computational cost must account for the neural network overhead. For ACE, the total cost takes the form of $aN^2 + bN^4$, where the $aN^2$ term arises from the neural network evaluations (e.g., CNN processing) across $N$ updates, and the $bN^4$ term originates from the determinant calculations. Because the neural network prefactor $a$ is relatively large, the asymptotic $\mathcal{O}(N^4)$ limit is not yet fully dominant for the finite system sizes studied here. This interplay results in the observed intermediate empirical scaling of roughly $\mathcal{O}(N^{2.19})$.

In contrast, SCALE is explicitly designed to overcome this bottleneck. By leveraging cached convolutional computations, the feature update cost becomes $\mathcal{O}(1)$. The locality of this ansatz ensures that the determinant can be updated iteratively using the Sherman-Morrison-Woodbury formula, reducing the cost from $\mathcal{O}(N^3)$ to $\mathcal{O}(N^2)$ for each update \cite{PhysRevB.40.506}. Consequently, the total theoretical complexity for an optimization step of $\mathcal{O}(N)$ updates is reduced to $\mathcal{O}(N^3)$. Analogous to the global case, the practical cost for SCALE is a combination of efficient local neural network updates and the iterative determinant updates, yielding a total cost model of $aN + bN^3$. Within our tested system sizes, the efficient $\mathcal{O}(N)$ feature updates heavily mitigate the overall operational cost, resulting in an effective empirical scaling of approximately $\mathcal{O}(N^{1.65})$.

Figure~\ref{fig:2} validates this analysis with numerical data, showing the scaling of cost for each optimization step for SCALE, ACE, and Transformer backflow \cite{gu2025solving}. As we can see from the figure, for the largest system size tested ($N = 512$), SCALE completes an optimization step in approximately 11 seconds. In comparison, the ACE architecture requires about 120 seconds, while the Transformer needs over 400 seconds, resulting in speedups of roughly $10\times$ and $40\times$, respectively. The runtime measurements demonstrate that SCALE maintains a significantly lower effective scaling exponent ($N^{1.65}$) compared to the ACE ($N^{2.19}$) and Transformer ($N^{2.90}$) baselines, indicating that the efficiency gain becomes increasingly dramatic for large-scale simulations.

\subsection{Attractive Hubbard model}

\begin{figure*}
\includegraphics[width=0.99\textwidth]{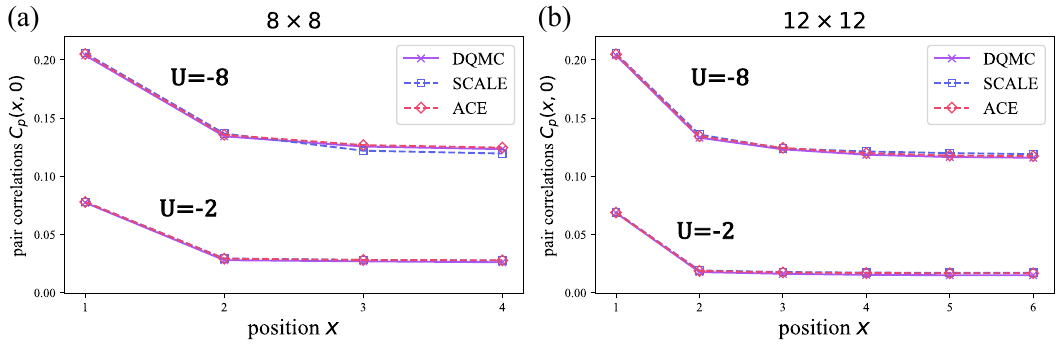}
\caption{\textbf{Pair correlations $C_p(x, 0)$ for the attractive Hubbard model.} The plots show results for (a) $8\times 8$ and (b) $12\times 12$ systems at weak ($U=-2$) and strong ($U=-8$) interaction strengths, with $t' = 0$. The results from our ACE and SCALE models perfectly align with the exact DQMC benchmarks, demonstrating their high accuracy. Statistical uncertainties from Monte Carlo sampling are smaller than the marker size in all panels.}
\label{fig:3}
\end{figure*}


To show the accuracy of our NQS architectures, we first apply them to the attractive Hubbard model with negative $U$. This model is a canonical benchmark for s-wave superconductivity, where the attractive on-site interaction induces the formation of Cooper pairs \cite{bardeen1957microscopic,PhysRevLett.62.1407}.

A standard method for quantifying this state in a particle-conserving system is through the pair-pair correlation function, $C_p(x, 0) = \langle\hat{\Delta}^\dagger(x)\hat{\Delta}(0)\rangle$, where $\hat{\Delta}(x) = c_{x\downarrow} c_{x\uparrow}$. The long-range behavior of $C_p(x, 0)$ as $x \to \infty$ serves as a direct proxy for the off-diagonal long-range order (ODLRO) \cite{yang1962concept} that characterizes the superconducting phase. Crucially, the attractive Hubbard model is free of the fermion sign problem in the spin-balanced sector, which allows DQMC simulations to provide numerically exact, unbiased results. The energies from ACE and SCALE agree perfectly well with the QMC results, and the comparison can be found in the appendix~\ref{tab:app1}.

In Figure~\ref{fig:3}, we plot pair correlations $C_p(x, 0)$ as a function of distance $x$ for two different system sizes and interaction strengths. We study both the weak coupling regime ($U=-2$) and the strong coupling regime ($U=-8$), choosing $t' = 0$ and $1/8$ hole doping. The results demonstrate outstanding accuracy for both ACE and SCALE. Across all systems and parameters tested, the correlations calculated by both SCALE and ACE are almost indistinguishable from the numerically exact DQMC benchmarks. The perfect alignment confirms that our methods can faithfully capture the complex quantum correlations of the s-wave superconducting ground state. 

The high accuracy of SCALE and ACE on the negative U model is supported by an analysis detailed in Appendix~\ref{app:ph}. We show that, through a particle-hole transformation, a BCS wavefunction can be explicitly mapped to our wave-function ansatz. This confirms that the generalized backflow transformation provides the necessary flexibility to accurately describe the superconducting ground state \cite{PhysRevB.78.041101} without requiring explicit pairing terms.


\subsection{Repulsive Hubbard model}

\begin{figure*}
\includegraphics[width=0.98\textwidth]{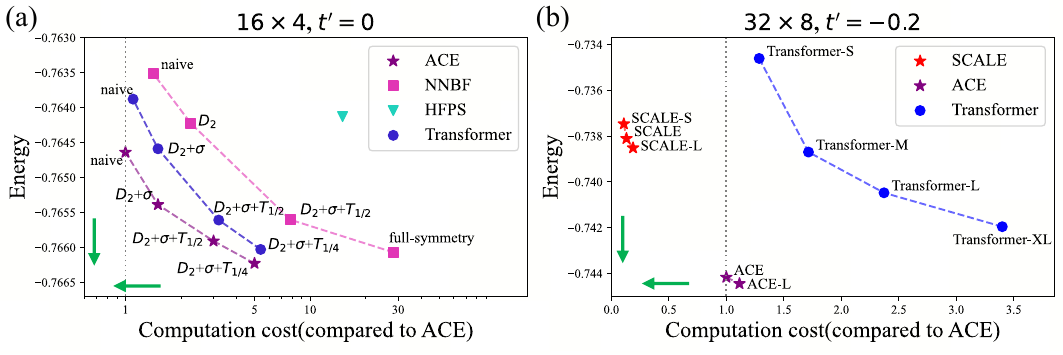}
\caption{\textbf{Accuracy vs. Cost trade-off on the $U=8$ Hubbard model.} The plots compare variational ground state energy against relative computational cost. (a) For the $16\times 4$ ($t'=0$) system, we analyze the effect of progressively enforcing symmetries on the ACE architecture (purple stars). Labels denote the cumulative addition of point group ($D_2$), spin-flip ($\sigma$), half-translation ($T_{1/2}$), and quarter-translation ($T_{1/4}$) symmetries. ACE achieves lower energies than Transformer~\cite{gu2025solving}, HFPS~\cite{chen2025neural}, and NNBF~\cite{loehr2025enhancingneuralnetworkbackflow} while maintaining a fraction of the computational cost of the full-symmetry NNBF. (b) On the larger $32\times 8$ ($t'=-0.2$) system, the ACE and SCALE variants establish a superior Pareto front, significantly outperforming Transformer-based models~\cite{gu2025solving} in both accuracy and computational efficiency.}
\label{fig:4}
\end{figure*}

\begin{table*}[t]
\centering
\begin{tabular}{c||c|c|c}
\toprule
Systems & $t'=0$, OBC & $ t'=0$, PBC & $t'=-0.2$, PBC \\
\midrule
PEPS \cite{liu2025} & -0.7260(2) &  & \\
HFPS+sym \cite{roth2025superconductivitytwodimensionalhubbardmodel} & & -0.7515(1) & -0.7360(1) \\
Tensor-Backflow \cite{liang2025investigating} & -0.7219 & -0.7509 & -0.7335\\
Tensor-Backflow+Lanczos \cite{liang2025investigating} & -0.7257 & -0.7552 & -0.7386\\
Transformer \cite{gu2025solving} & -0.7275 & -0.7563 & \\
SCALE & -0.7236 & -0.7529 & -0.7373\\
SCALE+GFMC & -0.7267 & -0.7560 & -0.7405\\
ACE & -0.7280 & -0.7573 & -0.7430\\
ACE+GFMC & \textbf{-0.7288}& \textbf{-0.7583} & \textbf{-0.7440}\\
\bottomrule
\end{tabular}
\caption{\textbf{Comparison of ground state energy for the $16\times 16$ Hubbard model.} Our SCALE and SCALE+GFMC are among the most accurate methods, while ACE and ACE+GFMC methods consistently find the lowest variational and projected energies, establishing new state-of-the-art results for this large system. For results where statistical errors are not shown, they are one or several orders of magnitude smaller than the last displayed significant digit.}
\label{tab:1}
\end{table*}

We now turn to the 2D repulsive Hubbard model, an iconic model for strongly correlated many-body physics \cite{J.Huardbard} and is believed to be relevant for Cuprate superconductors \cite{zhang1988effective}. We focus on the most challenging and intriguing regime with $U=8$, and hole doping $\delta=1/8$. We compare the ACE and SCALE results with several leading methods to demonstrate their superior accuracy and efficiency.

\subsubsection{Accuracy on $16 \times 4$ systems}

We focus first on a relatively small system with size $16\times 4$ under PBC. In Fig~\ref{fig:4}(a), we plot the ground state energy against the total computational cost, comparing the energies from ACE with previous state-of-the-art results. As it is well established that explicitly projecting a wavefunction onto the correct symmetry subspace can significantly lower the variational energy \cite{chen2025neural, loehr2025enhancingneuralnetworkbackflow, sharma2025comparing}, we also evaluate the energy in the fully symmetric sector to provide a fair comparison against other symmetrized methods.

Our standard ACE architecture, trained without explicit symmetries, achieves a ground-state energy of $-0.76464(1)$. This result is already more accurate than that of HFPS \cite{chen2025neural} while requiring only one-fifteenth of the cost. By directly optimizing the projected, symmetrized state reach a new state-of-the-art energy of $-0.76623(1)$. This surpasses the previous best result $-0.766073(6)$ from NNBF \cite{loehr2025enhancingneuralnetworkbackflow} with only one-sixth of the computational cost. Additionally, a GFMC calculation with the ACE trial wave-function gives an even lower energy of $-0.76647(2)$. Details of the symmetry projection process can be found in the Appendix~\ref{app:sym}.

\subsubsection{Accuracy on large systems}

We then study a large system with size $32 \times 8$ and including the next-nearest-neighboring hopping $t'=-0.2$, a challenging regime where the trade-off between variational accuracy and computational cost is critical. Our new results with SCALE and ACE and the comparison with the transformer backflow results \cite{gu2025solving} are summarized in Fig~\ref{fig:4}(b), where the ground state energy against computational cost is plotted. For the sake of simplicity, from here we report energies without symmetry projection. We want to emphasize that the energies can be further lowered by performing symmetry projections.

We further push the system size to as large as $32 \times 32$ (with $t'=0$), achieving a variational ground-state energy of $-0.7501$, and can be further reduced to $-0.7548$ using GFMC. The filled stripe phase is stabilized as shown in Appendix~\ref{appc}. This confirms the scalability of our method.

As Fig~\ref{fig:4}(b) shows, our ACE and SCALE results clearly establish a new, superior Pareto front, surpassing the previous state-of-the-art Transformer-based architectures~\cite{gu2025solving} by a large margin. The plot highlights two distinct advantages offered by our new architectures. SCALE is exceptionally efficient, being 10x faster than the small Transformer while achieving a better variational energy. ACE achieves a dramatically lower ground-state energy than any other method. Remarkably, it reaches this new level of accuracy while also being more computationally efficient than all the tested Transformer architectures. These results confirm that our proposed new architectures offer a drastically better trade-off between cost and accuracy. SCALE enables blazing-fast computations with high efficiency, whereas ACE redefines state-of-the-art accuracy without sacrificing practical runtime. Details of the configurations for each architecture are listed in Table \ref{tab:model}.

\subsubsection{Comparison with other leading methods}

In this subsection, we show a comparison of our results with other leading methods for a large-scale $16\times 16$ system. We consider systems under different boundary conditions, i.e., OBC and PBC, and systems with and without $t'$. As shown in Table~\ref{tab:1}, both SCALE and ACE demonstrate exceptional performance. SCALE is highly competitive, delivering accuracy on par with previous leading methods, including PEPS~\cite{liu2025}, HFPS~\cite{roth2025superconductivitytwodimensionalhubbardmodel}, Tensor-backflow~\cite{liang2025investigating}, and Transformer~\cite{gu2025solving}. In the meantime, ACE set new records in all studied cases. These are the lowest energies reported to date for these systems, confirming the state-of-the-art status of our method in large-scale simulation of strongly correlated quantum many-body systems. 

\subsubsection{Orientation of Stripe Order}

\begin{table*} 
	\centering

	\begin{tabular}{c||c|c} 
		\toprule
		Systems & $t'=0$, Filled Stripe ($\lambda=8$) & $t'=-0.2$, Half-Filled Stripe ($\lambda=4$)\\
		\midrule
		SCALE,hori & -0.7507 & -0.7371 \\
        SCALE,vert & -0.7507 & -0.7359 \\
        SCALE+GFMC,hori & -0.7546 & -0.7406\\
        SCALE+GFMC,vert & -0.7547 & -0.7397\\
		\bottomrule
	\end{tabular}
    \caption{\textbf{Energy of horizontal versus vertical stripes on the $32 \times 16$ Hubbard model with PBC.}
		The two systems shown are a filled stripe at $t' = 0$ and a half-filled stripe at $t' = -0.2$. The results from both SCALE and SCALE+GFMC show a clear energy preference for the horizontal stripe in the half-filled case, while there is no significant difference between the horizontal and vertical stripe energies within our accuracy for the filled stripe case with $t' =0$.}
    \label{tab:2} 
\end{table*}

In a previous work \cite{gu2025solving}, it was found that the half-filled stripe state for the Hubbard model with $t' = -0.2$ tends to align along the horizontal direction when the studied system has a rectangular geometry (i.e., the length $L_x$ is greater than the width $L_y$). In this context, a horizontal stripe refers to states where the rivers of holes running along the longer $x$-direction, whereas in a vertical stripe, the river of holes runs along the shorter $y$-direction. Ref.~\cite{gu2025solving} reported that the vertical stripe has a significantly higher energy. Their analysis showed that a horizontal stripe allows the doped holes to delocalize over a longer contiguous distance, which provides a more favorable gain in kinetic energy, while the potential energy difference between the two orientations remains small. It was proposed \cite{gu2025solving} to also study the preference of stripe orientation of the filled stripe in the pure Hubbard model without $t'$ in finite systems. However, such a comparison requires simulating much larger systems (e.g., a $32\times 16$ lattice for $1/8$ doping), which was computationally intractable using the transformer backflow with the resources available then. The superior efficiency of our new SCALE architecture makes these large-scale simulations feasible. We calculate the energies for both the horizontal and vertical filled stripe states for the $32 \times 16$ system, and the results are listed in Table \ref{tab:2}. For the $t'=-0.2$ case, we find similar results, i.e., the horizontal stripe has a lower energy. But for the $t'=0$ case, the energies for the horizontal and vertical stripe state are comparable within the accuracy of our method. A possible reason for this result is that the wavelength for the filled stripe at $1/8$ is already very large ($\lambda=16$ for spin density), so the kinetic energies of the doped hole in a hole river with length $16$ and $32$ have little difference.

\subsection{$t-J$ model}

\begin{figure*}
\includegraphics[width=0.6\textwidth]{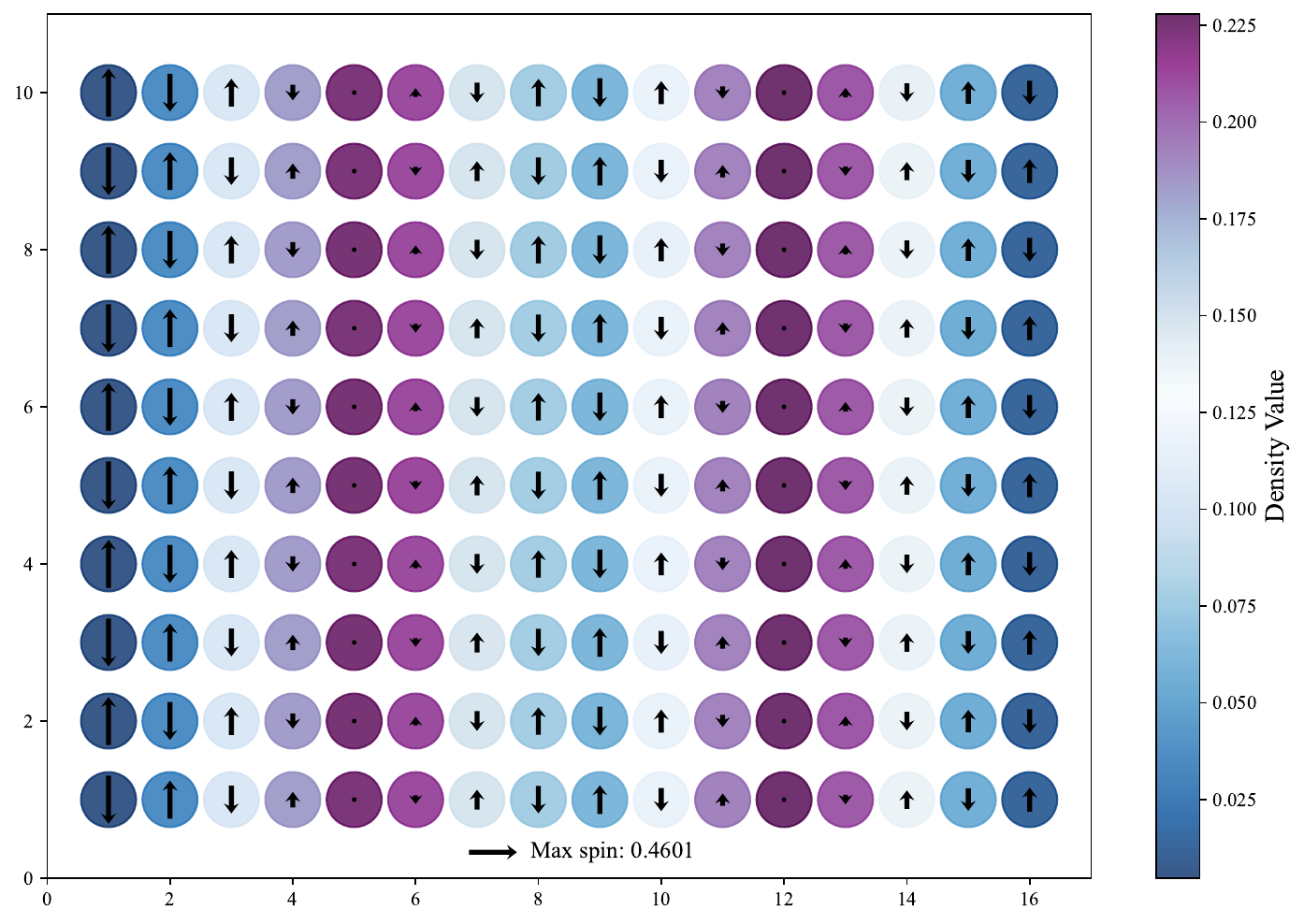}
\caption{\textbf{Ground state density and spin distribution for the $16 \times 10$ $t-J$ model with $J/t=1/3, t'/t=0.2, h_m=0.5$ at $1/8$ doping under cylindrical boundary conditions.} The color of the circles indicates the local hole density, and the arrows represent the magnitude and direction of the local spin density. We can see a clear filled stripe state from the results.}
\label{fig:5}
\end{figure*}

\begin{table*} 
	\centering
	\label{tab:3} 

	\begin{tabular}{c||c|c|c|c} 
		\toprule
		Systems & $32\times 4$ & $32\times 6$ & $16\times 8$ & $16\times 10$ \\
		\midrule
		DMRG & \textbf{-0.5071} & \textbf{-0.4994} & -0.5000 & -0.4961\\
		SCALE & -0.5025 & -0.4926 & -0.4974 & -0.4974\\
        SCALE+GFMC & -0.5044 & -0.4942 &  -0.4986 & -0.4985\\
        ACE & -0.5066 & -0.4964 & -0.5001 & -0.5001 \\
        ACE+GFMC & -0.5067 & -0.4967 & \textbf{-0.5003} & \textbf{-0.5003} \\
		\bottomrule
	\end{tabular}
    \caption{\textbf{The comparison of energies for $t-J$ model with $J/t=1/3, t'/t=0.2, h_m=0.5$ at $1/8$ doping.} For wider cylinders, SCALE and ACE achieve comparable and lower energies to DMRG with bond dimension $D=30000$ \cite{shen2025ground}.}
\end{table*}

We also apply our methods to the $t-J$ model, which is the effective Hamiltonian for the Hubbard model in the strong interaction limit.

The crucial feature of the $t-J$ model is the constraint of no double occupancy, which is enforced by the Gutzwiller projection operator, $\mathcal{P}_G = \prod_i (1 - n_{i\uparrow} n_{i\downarrow})$ as shown in Eq.~(\ref{ham-tj}). This operator projects the wavefunction onto the subspace of states without double occupancy. In our variational Monte Carlo framework, this constraint is implemented explicitly. The physical wavefunction amplitude is computed as $\Psi(\mathbf{n}) = \mathcal{P}_G \psi_\theta(\mathbf{n})$. In practice, this means that any proposed configuration $\mathbf{n}$ containing a doubly occupied site is immediately discarded by setting its amplitude to zero.

We follow the setup in a recent large-scale DMRG calculation in Ref.~\cite{shen2025ground} and set the parameters to $J/t = 1/3$, $t'/t = 0.2$, and a hole doping level of $\delta = 1/8$, which corresponds to electron doped situation with $t'/t=-0.2$ if the connection to the Hubbard model is considered. The simulations are performed on cylindrical boundary conditions. To help stabilize the expected order, an antiferromagnetic pinning field with strength $h_m = 0.5$ is applied to the first column of sites on the left open edge. Table~\ref{tab:3} presents our results, comparing SCALE and ACE (with and without GFMC projection) to highly accurate DMRG data (with bond dimension $D=30000$) for $32\times 4$, $32\times 6$, $16\times 8$, and $16 \times 10$ lattices. As expected, DMRG is more accurate in narrow cylinder systems with width $4$ and $6$. But when the width of the system is increased to $8$, the SCALE and ACE results are comparable to the DMRG energy. For an even wider cylinder with width $10$, the SCALE and ACE energies are much lower than the DMRG energy. 

It was found that $1/8$ doping is likely to lie in the boundary of the transition of the Neel anti-ferromagnetic phase with lower doping and a filled stripe phase with higher doping levels \cite{shen2025ground}. In Fig.~\ref{fig:5} we show the spin and hole density of the $16\times 10$ $t-J$ model, which shows a clear filled stripe order and is consistent with the conclusion in \cite{shen2025ground}, given that the ground state should be anti-ferromagnetic at very low dopings. We leave the determination of the precise phase boundary in this model for future investigations.

These results show that our new architectures are exceptionally accurate for the $t-J$ model, especially for larger (wider) system sizes. It was found in \cite{PhysRevB.102.041106,doi:10.1126/science.adh7691,shen2025ground,PhysRevLett.132.066002,jiang2025competitionchargedensitywavesuperconductingorders} that a large finite size effect exists in the Hubbard and $t-J$ model when the $t'$ term is included. So our new SCALE and ACE methods provide alternative accurate tools in determining the ground state of the Hubbard and $t'$ model in the thermodynamic limit.  

\section{Conclusion and Perspectives}

In this work, we have introduced a powerful and complementary pair of neural quantum state architectures, SCALE and ACE, which together establish a new Pareto frontier of the NQS for the simulation of strongly correlated many-fermion systems on lattice. This framework directly addresses the fundamental trade-off between computational efficiency and variational accuracy that has long defined the limits of the field. By providing two specialized tools, one for unprecedented scalability (SCALE) and one for extraordinary accuracy (ACE), our work offers a flexible and robust toolkit to address key challenges in condensed matter physics. 

The development of SCALE represents a significant algorithmic breakthrough. Its tailored convolutional design, combined with the fast low-rank update scheme, successfully breaks the $\mathcal{O}(N^4)$ scaling barrier of conventional backflow methods, achieving a much more favorable $\mathcal{O}(N^3)$ complexity. This asymptotic improvement, which translates to a practical speed-up exceeding $40\times$ for $16 \times 32$ system, is not merely an incremental gain. It fundamentally alters the landscape of what is computationally feasible. We demonstrated this directly by addressing an open question regarding the orientation of stripe order of the pure Hubbard model (with only nearest-neighboring hopping $t$) on a $32 \times 16$ lattice, a system size previously intractable for other NQS methods with high accuracy. Our findings suggest that, unlike partially filled stripes in the Hubbard model with next nearest-neighboring hoppings \cite{gu2025solving}, the horizontal and vertical filled stripes in the pure Hubbard model have comparable energy within the accuracy of our method. Moreover, with the friendly scaling of SCALE, we can reach the unprecedented $32 \times 32$ system for the pure Hubbard model. 

Furthermore, the high accuracy of SCALE, which remains competitive with global-range architectures like the Transformer-backflow \cite{gu2025solving}, challenges the prevailing assumption that mechanisms capturing long-range interactions at the input layer are a prerequisite for highly accurate ground-state simulations. Our results suggest that a deep yet local convolutional backflow is exceptionally effective at propagating information and capturing the complex, long-range correlation characteristic of the ground state for strongly correlated systems.

On the other end of the frontier, ACE establishes a new high-water mark for variational accuracy. By replacing the efficient sparse layer with a deep convolutional stack, ACE maximizes expressive power. As shown in Figure \ref{fig:4} and Table \ref{tab:1}, it consistently sets new state-of-the-art energy benchmarks, decisively surpassing the most accurate methods recently reported, including Transformer-based models \cite{gu2025solving} and enhanced backflows \cite{loehr2025enhancingneuralnetworkbackflow}. Crucially, ACE achieves this new level of accuracy while being significantly more computationally efficient than its competitors, requiring, for example, only one-sixth of the runtime of NNBF for a more accurate result for a system with size $16\times 4$. This combination of precision and relative efficiency is vital for the study of strongly correlated systems, where the central challenge, as noted in our introduction, is often the resolution of competing states with nearly degenerate energies \cite{qin2022hubbard}. ACE provides a powerful and reliable new tool for this exact purpose.

In summary, we have introduced a unified (SCALE/ACE) framework that redefines the state-of-the-art in variational simulations of strongly correlated lattice fermions, providing a powerful and scalable path forward. The efficiency of SCALE, with its $\mathcal{O}(N^3)$ scaling, enables systematic, large-scale studies of finite-size effects, allowing for more reliable extrapolations to the thermodynamic limit. Simultaneously, the extraordinary accuracy of ACE makes it an ideal instrument for a definitive search for elusive phases, such as d-wave superconductivity, in the repulsive Hubbard and $t-J$ models. Future work can leverage this platform to explore key physical problems, such as the microscopic mechanism of high-Tc cuprate superconductors. Extending these methods beyond ground-state calculations presents another promising research direction.

\begin{acknowledgments}
We thank ByteDance Seed for inspiration and encouragement, and Hang Li for his guidance and support.
We thank Yuan-Yao He and Hao Du for providing DQMC data.
We thank Yang Shen for providing DMRG data.
L.W. is supported by National Science and Technology Major Project (2022ZD0114902) and National Science Foundation of China (NSFC92470123, NSFC62276005). M.Q. acknowledges
the support from the National Key Research and Development Program of MOST of China (2022YFA1405400),
the National Natural Science Foundation of China (Grant
No. 12274290 and No. 12522406), and the Innovation Program for Quantum Science and Technology
(2021ZD0301902)
\end{acknowledgments}

\appendix

\section{Representability of Pairing Wavefunctions}

\label{app:ph}

A general mean-field wavefunction can be written as the Thouless state, $\ket{\psi} = \exp \left( \frac{1}{2} \sum_{p, q} F_{pq} \hat{c}^\dagger_p c^\dagger_{q} \right) \ket{0}$, where $p, q$ runs over all orbital and spin indices, and $F$ is an anti-symmetric matrix. The BCS state, in particular, can be viewed as a special instance. The ground state of the attractive Hubbard model, despite its many-body nature, is believed to be close to such mean-field wavefunctions.

When benchmarking our method on the attractive Hubbard model, we use a particle-hole transformation (PHT) defined by $U = \prod_{i=1}^N \left(c^\dagger_{i\downarrow} + (-1)^i c_{i\downarrow}\right)$, where $N$ is the number of sites, and $(-1)^i$ gives opposite signs to different sublattices. Under this PHT, we have
$c_{i\uparrow} \to c_{i\uparrow}$, $c_{i\downarrow} \to (-1)^i c_{i\downarrow}^\dagger$, and the Hubbard Hamiltonian transforms as $\hat{H} \to \hat{H}_{\mathrm{PH}} = -t \sum_{\braket{i, j}, \sigma} \left(c^\dagger_{i\sigma} c_{j\sigma} + \mathrm{h.c.}\right) + U \sum_i n_{i\uparrow} (1 - n_{i\downarrow})$. 

Specifically, a BCS wavefunction 
\begin{math}
    \ket{\psi_{\mathrm{BCS}}} \propto \exp \left( \sum_{i, j} F_{ij} \hat{c}^\dagger_{i\uparrow} c^\dagger_{j \downarrow} \right) \ket{0},
\end{math}
is transformed to a particle-number-conserved one 
\begin{equation}
    \ket{\Psi_{\mathrm{BCS}}} \propto \exp \left( \sum_{i, j=1}^N (-1)^j F_{i, j} c^\dagger_{i\uparrow} c_{j\downarrow} \right) \prod^{N}_{l=1} c^\dagger_{l \downarrow} \ket{0},    
\end{equation}
this is a Slater determinant, as can be verified using the identity
\begin{equation}
    \begin{split}
        &\exp\left( \sum_{i, j=1}^N M_{i, j} c^\dagger_{i\uparrow} c_{j\downarrow}\right) c^\dagger_{l\downarrow} \exp\left( - \sum_{i,j=1}^N M_{i, j} c^\dagger_{i\uparrow} c_{j\downarrow}\right) \\ 
        & \quad = c^\dagger_{l\downarrow} + \sum_i c^\dagger_{i \uparrow} M_{i l}.
    \end{split}
\end{equation}

By taking $M_{ij} = (-1)^j F_{i, j}$, we have
\begin{equation}
    \begin{split}
        \braket{\bm n | \Psi_{\mathrm{BCS}}} &\propto \bra{\bm n } \prod_{l=1}^N \left(c^\dagger_{l\downarrow} + \sum_i c^\dagger_{i \uparrow} M_{i l} \right) \ket{0} \\
        & = \det\left[ \bm n \star
        \begin{pmatrix}
        M \\
        I
        \end{pmatrix}
        \right]
    \end{split},
\end{equation}
where $\bm n \star$ denotes selecting $N$ rows of the subsequent matrix based on the occupied orbitals of $\bm n$. This form is well within the representability of our NQS ansatz, which motivates us to simulate the particle-hole transformed Hamiltonian $\hat{H}_{\mathrm{PH}}$ instead, with the intuition that the s-wave pairing presented in this model is better captured.

\section{Symmetry Projection}
\label{app:sym}

\begin{figure}
\includegraphics[width=0.45\textwidth]{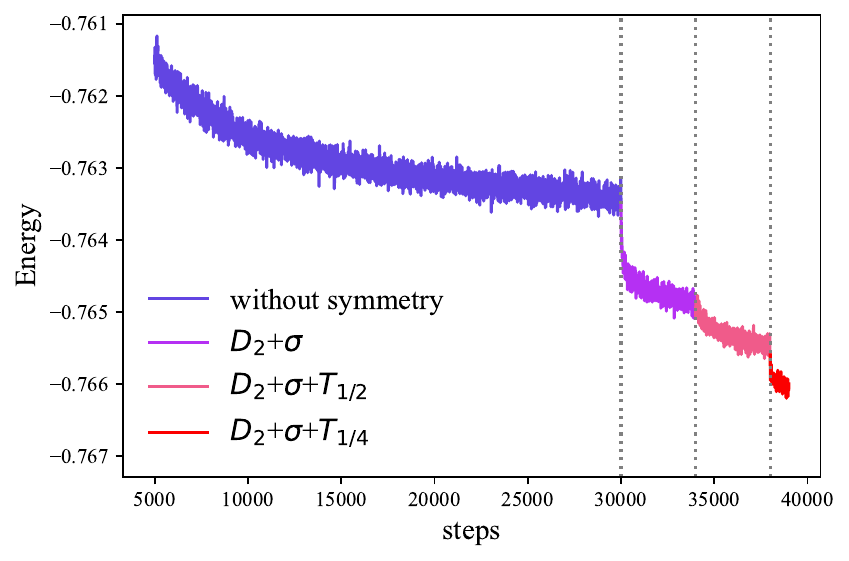}
\caption{\textbf{Variational energy of ACE as a function of optimization steps.} The system is a Hubbard model with size $16\times 4$ and $t'=0$. We can see that symmetry projection greatly increases the variational accuracy.}
\label{fig:6}
\end{figure}

The symmetrized wavefunction, $\psi _{\text{sym}}$, is obtained by projecting the initial wavefunction:
\begin{equation}
\psi _{\text{sym}} = \sum _g \Pi (\mathbf{n}, g)\chi_g \psi (g\mathbf{n}),
\end{equation}
where $g \in G$ is an operation from the symmetry group, $\Pi (\mathbf{n}, g) = \pm 1$ is the fermionic sign from permutations and $\chi_g$ is the character of the symmetry representation. 

Our procedure involves two stages. In the first stage, we perform $3\times 10^4$ initial steps without enforcing any symmetry. The resulting state is then projected onto the fully symmetric state, a step with essentially no computational cost. This baseline projection corresponds to ``naive'' in Figure~\ref{fig:4}(a).

In the second stage, we gradually enforce symmetry following the procedure in Ref.\cite{viteritti2025transformer} through a sequence of training steps as shown in Figure~\ref{fig:6}. First, we restore $D_2$ and spin inversion symmetry and train for $4 \times 10^3$ steps. Next, we additionally restore the half-translation symmetries and train for another $4 \times 10^3$ steps. Finally, we restore the 1/4-translational symmetries and train for $10^3$ steps. After this sequential training, the state is once again projected onto the fully symmetric state at essentially no cost.

\section{Pinning Field}

To avoid convergence to local minima during optimization, temporary pinning fields are applied following Ref.\cite{liu2025, gu2025solving}. Different pinning fields are imposed to get desired patterns for a simple backflow network. Next, the pattern is transferred to SCALE or ACE using a pretraining method. The final optimization with MARCH is performed with the pinning field removed, allowing the wave-function to relax and find the true ground state. With this method, we can stabilize vertical or horizontal stripe states with different fillings or wave-lengths.

\section{Additional Data}

\label{appc}

\begin{figure*}
\includegraphics[width=0.95\textwidth]{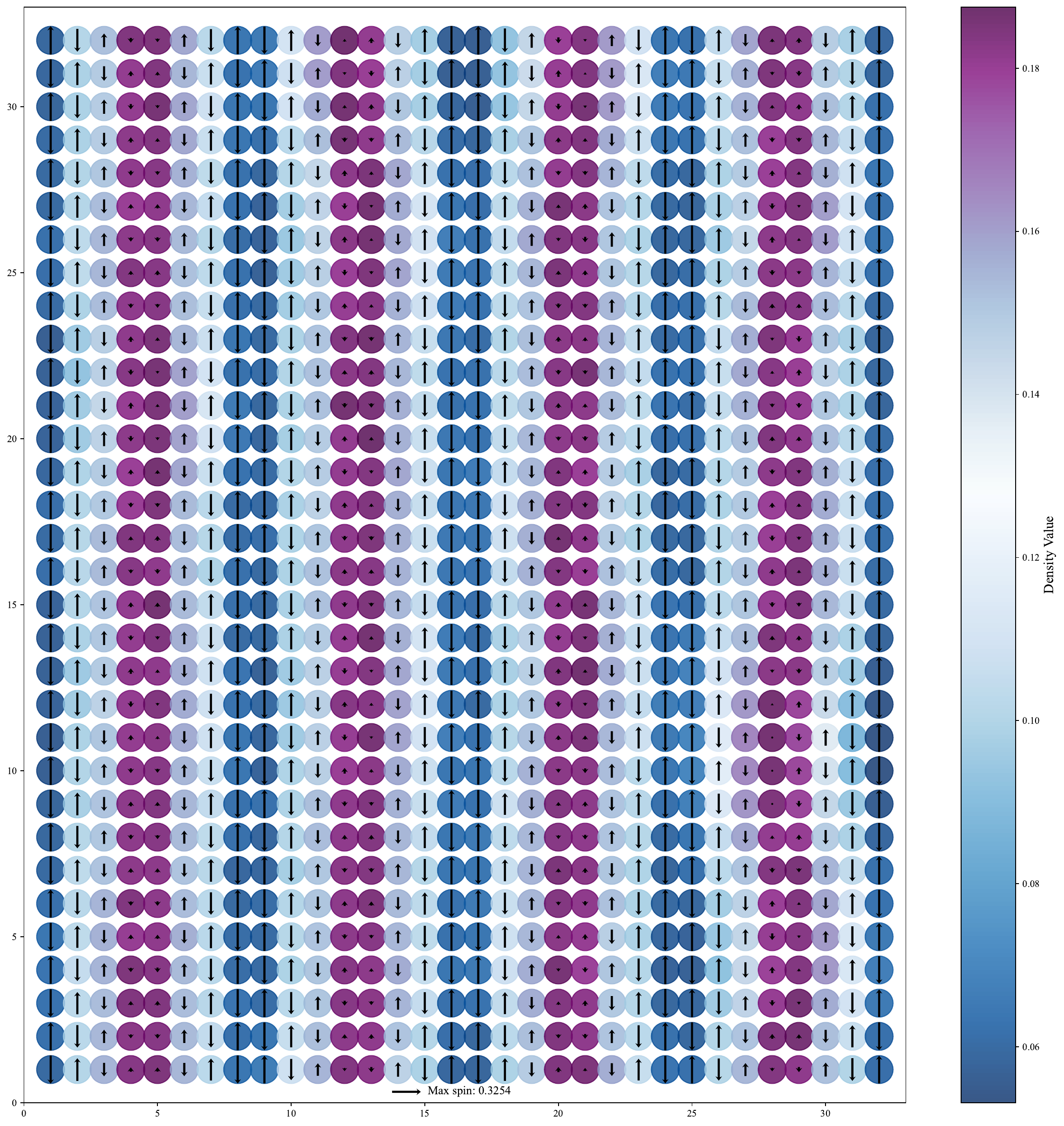}
\caption{\textbf{Ground state density and spin distribution for the $32 \times 32$ Hubbard model with $t'=0$ at $1/8$ doping under PBC.} The color of the circles indicates the local hole density, and the arrows represent the magnitude and direction of the local spin density. We can see a clear filled stripe state from the results.}
\label{fig:7}
\end{figure*}

We provide additional benchmark results for the attractive Hubbard model in Table~\ref{tab:app1}. The table compares the ground state energies from our models (SCALE, ACE, and their GFMC variants) against established DQMC benchmarks for various system sizes and interaction strengths. We also provide the ground state charge and spin density for $32\times 32$ Hubbard model in Figure~\ref{fig:7}.

\begin{table*} 
	\centering
	\begin{tabular}{c||c|c|c|c} 
		\toprule
		Systems & $8\times 8, U=-2$ & $8\times 8, U=-8$ & $12\times 12, U=-2$ & $12\times 12, U=-8$ \\
		\midrule
		DQMC & -2.0343 & -4.017(1) & -2.0416 & -4.016(1)\\
		SCALE & -2.0337 & -4.0161 & -2.0404 & -4.0141\\
        SCALE+GFMC & -2.0338 & -4.0167 & -2.0406 & -4.0158\\
        ACE & -2.0338 & -4.0165 & -2.0405 & -4.0159\\
        ACE+GFMC & -2.0339 & -4.0168 & -2.0407 & -4.0161\\
		\bottomrule
	\end{tabular}
    \caption{\textbf{Comparison of the energies in the attractive Hubbard model.}
		The energies from SCALE and ACE agree perfectly well with DQMC.}
    \label{tab:app1}
\end{table*}

\section{Hyperparameter}

For full reproducibility, this section outlines the specific parameters used in our experiments. We provide the settings for the MARCH optimizer, followed by the architectural details for the SCALE and ACE model variants.

\begin{table*}[t!] 
	\centering
	\label{tab:opt} 
	\begin{tabular}{c || c | c}
		\toprule
		Hyperparameter & ACE & SCALE \\
        \midrule
		Learning rate schedule ($t$) & $10^{-1}(1 + \max(t \!-\! 1000, 0)/4000)^{-1}$ & $5 \times 10^{-1}(1 + \max(t \!-\! 1000, 0)/8000)^{-1}$ \\
		\midrule
		Optimizer momentum ($\mu, \beta$) & \multicolumn{2}{c}{0.95, 0.995} \\
		Damping ($\lambda$) & \multicolumn{2}{c}{0.001} \\
		Batch size & \multicolumn{2}{c}{$\text{clip}(32 L_x \times L_y, 4096, 8192)$} \\
		Local energy clipping & \multicolumn{2}{c}{5.0} \\
		MCMC steps per sample & \multicolumn{2}{c}{$2.5 L_x \times L_y$} \\
		Total optimization steps & \multicolumn{2}{c}{50000} \\
		\bottomrule
	\end{tabular}
    \caption{\textbf{Hyperparameters for the MARCH optimizer.} Shared parameters apply to both models, while model-specific settings are listed in their respective columns.}
\end{table*}

\newcommand{\kernelS}{
    \begin{tikzpicture}[scale=0.10, baseline={(0,0.05)}] 
        \foreach \x in {0,1,2} {
            \foreach \y in {0,1,2} {
                \fill[black!80] (\x,\y) rectangle ++(1,1);
                \draw[white, line width=0.5pt] (\x,\y) rectangle ++(1,1);
            }
        }
    \end{tikzpicture}
}

\newcommand{\kernelL}{
    \begin{tikzpicture}[scale=0.10, baseline={(0,0.15)}]
        \foreach \x in {0,1,2} {
            \foreach \y in {0,1,2} {
                \fill[black!80] (\x+1,\y+1) rectangle ++(1,1); 
                \draw[white, line width=0.5pt] (\x+1,\y+1) rectangle ++(1,1);
            }
        }
        \foreach \p in {(2,4), (2,0), (0,2), (4,2)} {
            \fill[black!80] \p rectangle ++(1,1);
            \draw[white, line width=0.5pt] \p rectangle ++(1,1);
        }
    \end{tikzpicture}
}

\begin{table*}[t!]
	\centering
	\label{tab:model}
	\begin{tabular}{c || c |c |c| c | c} 
		\toprule
		Model & $h_{\text{conv}}$ & $h_{\text{MLP}}$ & $L_{\text{conv}}$ & $L_{\text{MLP}}$ & Kernel Shape \\
		\midrule
		SCALE-S & 256 & 512 & 1 & 4 & \kernelS \\
		SCALE   & 512 & 512 & 1 & 4 & \kernelS \\
		SCALE-L & 384 & 512 & 1 & 4 & \kernelL \\
        SCALE-XL & 640 & 1280 & 1 & 4 & \kernelS \\
		\midrule
		ACE     & 144 & 256 & 16 & 2 & \kernelS \\
		ACE-L   & 192 & 256 & 16 & 2 & \kernelS \\
		\bottomrule
	\end{tabular}
    \caption{\textbf{Model hyperparameters for SCALE and ACE variants.} 
    We list the details of the convolutional hidden dimension ($h_{\text{conv}}$), 
    MLP hidden dimension ($h_{\text{MLP}}$), number of convolutional layers 
    ($L_{\text{conv}}$), and number of MLP layers ($L_{\text{MLP}}$). 
    Note that SCALE-L uses a convolutional filter with a neighborhood range of Euclidean distance $L_2 \le 2$, as opposed to the $L_2 \le \sqrt{2}$ range used in other models. The SCALE-XL hyperparameters are used for $32\times 32$ system size only.
	}
\end{table*}







\bibliography{main}

@article{qin2022hubbard,
  title={The Hubbard model: A computational perspective},
  author={Qin, Mingpu and Sch{\"a}fer, Thomas and Andergassen, Sabine and Corboz, Philippe and Gull, Emanuel},
  journal={Annual Review of Condensed Matter Physics},
  volume={13},
  number={1},
  pages={275--302},
  year={2022},
  publisher={Annual Reviews}
}

@article{arovas2022hubbard,
  title={The hubbard model},
  author={Arovas, Daniel P and Berg, Erez and Kivelson, Steven A and Raghu, Srinivas},
  journal={Annual review of condensed matter physics},
  volume={13},
  number={1},
  pages={239--274},
  year={2022},
  publisher={Annual Reviews}
}

@article{leblanc2015solutions,
  title={Solutions of the two-dimensional Hubbard model: Benchmarks and results from a wide range of numerical algorithms},
  author={LeBlanc, James PF and Antipov, Andrey E and Becca, Federico and Bulik, Ireneusz W and Chan, Garnet Kin-Lic and Chung, Chia-Min and Deng, Youjin and Ferrero, Michel and Henderson, Thomas M and Jim{\'e}nez-Hoyos, Carlos A and others},
  journal={Physical Review X},
  volume={5},
  number={4},
  pages={041041},
  year={2015},
  publisher={APS}
}

@article{blankenbecler1981monte,
  title = {Monte Carlo calculations of coupled boson-fermion systems. I},
  author = {Blankenbecler, R. and Scalapino, D. J. and Sugar, R. L.},
  journal = {Phys. Rev. D},
  volume = {24},
  issue = {8},
  pages = {2278--2286},
  numpages = {0},
  year = {1981},
  month = {Oct},
  publisher = {American Physical Society},
  doi = {10.1103/PhysRevD.24.2278},
  url = {https://link.aps.org/doi/10.1103/PhysRevD.24.2278}
}

@article{PhysRevB.28.4059,
  title = {Discrete Hubbard-Stratonovich transformation for fermion lattice models},
  author = {Hirsch, J. E.},
  journal = {Phys. Rev. B},
  volume = {28},
  issue = {7},
  pages = {4059--4061},
  numpages = {0},
  year = {1983},
  month = {Oct},
  publisher = {American Physical Society},
  doi = {10.1103/PhysRevB.28.4059},
  url = {https://link.aps.org/doi/10.1103/PhysRevB.28.4059}
}

@article{Hirsch1985two,
  title = {Two-dimensional Hubbard model: Numerical simulation study},
  author = {Hirsch, J. E.},
  journal = {Phys. Rev. B},
  volume = {31},
  issue = {7},
  pages = {4403--4419},
  numpages = {0},
  year = {1985},
  month = {Apr},
  publisher = {American Physical Society},
  doi = {10.1103/PhysRevB.31.4403},
  url = {https://link.aps.org/doi/10.1103/PhysRevB.31.4403}
}

@article{PhysRevLett.69.2863,
  title = {Density matrix formulation for quantum renormalization groups},
  author = {White, Steven R.},
  journal = {Phys. Rev. Lett.},
  volume = {69},
  issue = {19},
  pages = {2863--2866},
  numpages = {0},
  year = {1992},
  month = {Nov},
  publisher = {American Physical Society},
  doi = {10.1103/PhysRevLett.69.2863},
  url = {https://link.aps.org/doi/10.1103/PhysRevLett.69.2863}
}

@article{orus_practical_2014,
	title = {A practical introduction to tensor networks: {Matrix} product states and projected entangled pair states},
	volume = {349},
	issn = {0003-4916},
	url = {https://www.sciencedirect.com/science/article/pii/S0003491614001596},
	doi = {https://doi.org/10.1016/j.aop.2014.06.013},
	journal = {Annals of Physics},
	author = {Orús, Román},
	year = {2014},
	pages = {117--158},
}

@article{Cirac2021rmp,
  title = {Matrix product states and projected entangled pair states: Concepts, symmetries, theorems},
  author = {Cirac, J. Ignacio and P\'erez-Garc\'{\i}a, David and Schuch, Norbert and Verstraete, Frank},
  journal = {Rev. Mod. Phys.},
  volume = {93},
  issue = {4},
  pages = {045003},
  numpages = {65},
  year = {2021},
  month = {Dec},
  publisher = {American Physical Society},
  doi = {10.1103/RevModPhys.93.045003},
  url = {https://link.aps.org/doi/10.1103/RevModPhys.93.045003}
}

@misc{verstraete2004,
      title={Renormalization algorithms for Quantum-Many Body Systems in two and higher dimensions}, 
      author={F. Verstraete and J. I. Cirac},
      year={2004},
      eprint={cond-mat/0407066},
      archivePrefix={arXiv},
      primaryClass={cond-mat.str-el},
      url={https://arxiv.org/abs/cond-mat/0407066}, 
}

@article{carleo2017solving,
  title={Solving the quantum many-body problem with artificial neural networks},
  author={Carleo, Giuseppe and Troyer, Matthias},
  journal={Science},
  volume={355},
  number={6325},
  pages={602--606},
  year={2017},
  publisher={American Association for the Advancement of Science}
}

@article{luo2019backflow,
  title={Backflow transformations via neural networks for quantum many-body wave functions},
  author={Luo, Di and Clark, Bryan K},
  journal={Physical review letters},
  volume={122},
  number={22},
  pages={226401},
  year={2019},
  publisher={APS}
}

@article{robledo2022fermionic,
  title={Fermionic wave functions from neural-network constrained hidden states},
  author={Robledo Moreno, Javier and Carleo, Giuseppe and Georges, Antoine and Stokes, James},
  journal={Proceedings of the National Academy of Sciences},
  volume={119},
  number={32},
  pages={e2122059119},
  year={2022},
  publisher={National Academy of Sciences}
}

@article{gu2025solving,
  title={Solving the Hubbard model with Neural Quantum States},
  author={Gu, Yuntian and Li, Wenrui and Lin, Heng and Zhan, Bo and Li, Ruichen and Huang, Yifei and He, Di and Wu, Yantao and Xiang, Tao and Qin, Mingpu and others},
  journal={arXiv preprint arXiv:2507.02644},
  year={2025}
}

@article{liang2025investigating,
  title={Investigating the Fermi-Hubbard model by the Tensor-Backflow method},
  author={Liang, Xiao},
  journal={arXiv preprint arXiv:2507.01856},
  year={2025}
}

@article{chen2025neural,
  title={Neural Network-Augmented Pfaffian Wave-functions for Scalable Simulations of Interacting Fermions},
  author={Chen, Ao and Wan, Zhou-Quan and Sengupta, Anirvan and Georges, Antoine and Roth, Christopher},
  journal={arXiv preprint arXiv:2507.10705},
  year={2025}
}

@misc{loehr2025enhancingneuralnetworkbackflow,
      title={Enhancing Neural Network Backflow}, 
      author={Kieran Loehr and Bryan K. Clark},
      year={2025},
      eprint={2510.26906},
      archivePrefix={arXiv},
      primaryClass={cond-mat.str-el},
      url={https://arxiv.org/abs/2510.26906}, 
}

@article{liu2025,
  title = {Accurate Simulation of the Hubbard Model with Finite Fermionic Projected Entangled Pair States},
  author = {Liu, Wen-Yuan and Zhai, Huanchen and Peng, Ruojing and Gu, Zheng-Cheng and Chan, Garnet Kin-Lic},
  journal = {Phys. Rev. Lett.},
  volume = {134},
  issue = {25},
  pages = {256502},
  numpages = {8},
  year = {2025},
  month = {Jun},
  publisher = {American Physical Society},
  doi = {10.1103/r4q9-4yvj},
  url = {https://link.aps.org/doi/10.1103/r4q9-4yvj}
}

@article{shen2025ground,
  title={The ground state of electron-doped t- t'- J model on cylinders: an investigation of finite size and boundary condition effects},
  author={Shen, Yang and Qian, Xiangjian and Qin, Mingpu},
  journal={Chinese Physics B},
  year={2025}
}

@article{elfwing2018sigmoid,
  title={Sigmoid-weighted linear units for neural network function approximation in reinforcement learning},
  author={Elfwing, Stefan and Uchibe, Eiji and Doya, Kenji},
  journal={Neural networks},
  volume={107},
  pages={3--11},
  year={2018},
  publisher={Elsevier}
}

@article{scherbela2025accurate,
  title={Accurate Ab-initio Neural-network Solutions to Large-Scale Electronic Structure Problems},
  author={Scherbela, Michael and Gao, Nicholas and Grohs, Philipp and G{\"u}nnemann, Stephan},
  journal={arXiv preprint arXiv:2504.06087},
  year={2025}
}

@article{PhysRevB.41.4552,
  title = {Ground-state correlations of quantum antiferromagnets: A Green-function Monte Carlo study},
  author = {Trivedi, Nandini and Ceperley, D. M.},
  journal = {Phys. Rev. B},
  volume = {41},
  issue = {7},
  pages = {4552--4569},
  numpages = {0},
  year = {1990},
  month = {Mar},
  publisher = {American Physical Society},
  doi = {10.1103/PhysRevB.41.4552},
  url = {https://link.aps.org/doi/10.1103/PhysRevB.41.4552}
}

@article{PhysRevB.51.13039,
  title = {Proof for an upper bound in fixed-node Monte Carlo for lattice fermions},
  author = {ten Haaf, D. F. B. and van Bemmel, H. J. M. and van Leeuwen, J. M. J. and van Saarloos, W. and Ceperley, D. M.},
  journal = {Phys. Rev. B},
  volume = {51},
  issue = {19},
  pages = {13039--13045},
  numpages = {0},
  year = {1995},
  month = {May},
  publisher = {American Physical Society},
  doi = {10.1103/PhysRevB.51.13039},
  url = {https://link.aps.org/doi/10.1103/PhysRevB.51.13039}
}

@article{J.Huardbard,
 URL = {http://www.jstor.org/stable/2414761},
 author = {J. Hubbard},
 journal = {Proc. R. Soc. Lond. A},
 number = {1365},
 pages = {238--257},
 publisher = {Royal Society},
 title = {{Electron Correlations in Narrow Energy Bands}},
 volume = {276},
 year = {1963},
 doi = {10.1098/rspa.1963.0204}
}

@article{zhang1988effective,
  title={Effective Hamiltonian for the superconducting Cu oxides},
  author={Zhang, FC and Rice, TM},
  journal={Physical Review B},
  volume={37},
  number={7},
  pages={3759},
  year={1988},
  publisher={APS}
}

@article{bardeen1957microscopic,
  title={Microscopic theory of superconductivity},
  author={Bardeen, John and Cooper, Leon N and Schrieffer, J Robert},
  journal={Physical Review},
  volume={106},
  number={1},
  pages={162},
  year={1957},
  publisher={APS}
}

@article{yang1962concept,
  title = {Concept of Off-Diagonal Long-Range Order and the Quantum Phases of Liquid He and of Superconductors},
  author = {Yang, C. N.},
  journal = {Rev. Mod. Phys.},
  volume = {34},
  issue = {4},
  pages = {694--704},
  numpages = {0},
  year = {1962},
  month = {Oct},
  publisher = {American Physical Society},
  doi = {10.1103/RevModPhys.34.694},
  url = {https://link.aps.org/doi/10.1103/RevModPhys.34.694}
}

@article{sharma2025comparing,
  title={Comparing Symmetrized Determinant Neural Quantum States for the Hubbard Model},
  author={Sharma, Louis and Shokry, Ahmedeo and Nutakki, Rajah and Simard, Olivier and Ferrero, Michel and Vicentini, Filippo},
  journal={arXiv preprint arXiv:2510.11710},
  year={2025}
}

@misc{roth2025superconductivitytwodimensionalhubbardmodel,
      title={Superconductivity in the two-dimensional Hubbard model revealed by neural quantum states}, 
      author={Christopher Roth and Ao Chen and Anirvan Sengupta and Antoine Georges},
      year={2025},
      eprint={2511.07566},
      archivePrefix={arXiv},
      primaryClass={cond-mat.supr-con},
      url={https://arxiv.org/abs/2511.07566}, 
}

@article{viteritti2025transformer,
  title={Transformer wave function for two dimensional frustrated magnets: Emergence of a spin-liquid phase in the Shastry-Sutherland model},
  author={Viteritti, Luciano Loris and Rende, Riccardo and Parola, Alberto and Goldt, Sebastian and Becca, Federico},
  journal={Physical Review B},
  volume={111},
  number={13},
  pages={134411},
  year={2025},
  publisher={APS}
}

@article{ba2016layer,
  title={Layer normalization},
  author={Ba, Jimmy Lei and Kiros, Jamie Ryan and Hinton, Geoffrey E},
  journal={arXiv preprint arXiv:1607.06450},
  year={2016}
}

@article{PhysRevLett.20.1445,
  title = {Absence of Mott Transition in an Exact Solution of the Short-Range, One-Band Model in One Dimension},
  author = {Lieb, Elliott H. and Wu, F. Y.},
  journal = {Phys. Rev. Lett.},
  volume = {20},
  issue = {25},
  pages = {1445--1448},
  numpages = {0},
  year = {1968},
  month = {Jun},
  publisher = {American Physical Society},
  doi = {10.1103/PhysRevLett.20.1445},
  url = {https://link.aps.org/doi/10.1103/PhysRevLett.20.1445}
}

@article{RevModPhys.87.457,
  title = {Colloquium: Theory of intertwined orders in high temperature superconductors},
  author = {Fradkin, Eduardo and Kivelson, Steven A. and Tranquada, John M.},
  journal = {Rev. Mod. Phys.},
  volume = {87},
  issue = {2},
  pages = {457--482},
  numpages = {26},
  year = {2015},
  month = {May},
  publisher = {American Physical Society},
  doi = {10.1103/RevModPhys.87.457},
  url = {https://link.aps.org/doi/10.1103/RevModPhys.87.457}
}

@article{PhysRevB.102.041106,
  title = {Plaquette versus ordinary $d$-wave pairing in the ${t}^{\ensuremath{'}}$-Hubbard model on a width-4 cylinder},
  author = {Chung, Chia-Min and Qin, Mingpu and Zhang, Shiwei and Schollw\"ock, Ulrich and White, Steven R.},
  collaboration = {The Simons Collaboration on the Many-Electron Problem},
  journal = {Phys. Rev. B},
  volume = {102},
  issue = {4},
  pages = {041106},
  numpages = {6},
  year = {2020},
  month = {Jul},
  publisher = {American Physical Society},
  doi = {10.1103/PhysRevB.102.041106},
  url = {https://link.aps.org/doi/10.1103/PhysRevB.102.041106}
}

@article{
doi:10.1126/science.adh7691,
author = {Hao Xu  and Chia-Min Chung  and Mingpu Qin  and Ulrich Schollwöck  and Steven R. White  and Shiwei Zhang },
title = {Coexistence of superconductivity with partially filled stripes in the Hubbard model},
journal = {Science},
volume = {384},
number = {6696},
pages = {eadh7691},
year = {2024},
doi = {10.1126/science.adh7691},
URL = {https://www.science.org/doi/abs/10.1126/science.adh7691},
eprint = {https://www.science.org/doi/pdf/10.1126/science.adh7691}}

@article{PhysRevLett.75.3537,
  title = {Thermodynamic Limit of Density Matrix Renormalization},
  author = {\"Ostlund, Stellan and Rommer, Stefan},
  journal = {Phys. Rev. Lett.},
  volume = {75},
  issue = {19},
  pages = {3537--3540},
  numpages = {0},
  year = {1995},
  month = {Nov},
  publisher = {American Physical Society},
  doi = {10.1103/PhysRevLett.75.3537},
  url = {https://link.aps.org/doi/10.1103/PhysRevLett.75.3537}
}

@article{SCHOLLWOCK201196,
title = {The density-matrix renormalization group in the age of matrix product states},
journal = {Annals of Physics},
volume = {326},
number = {1},
pages = {96-192},
year = {2011},
note = {January 2011 Special Issue},
issn = {0003-4916},
doi = {https://doi.org/10.1016/j.aop.2010.09.012},
url = {https://www.sciencedirect.com/science/article/pii/S0003491610001752},
author = {Ulrich Schollwöck}
}

@book{xiang2023density,
  url={https://books.google.com/books?hl=en&lr=&id=E5fxEAAAQBAJ&oi=fnd&pg=PP1&ots=Hqdq-ApAx0&sig=xi-IvwDkPLk7CTWPcB_d5zRtFZk},
  title={{Density Matrix and Tensor Network Renormalization}},
  author={Xiang, Tao},
  year={2023},
  publisher={Cambridge University Press}
}

@article{PhysRevLett.123.136402,
  title = {Reaching the Continuum Limit in Finite-Temperature Ab Initio Field-Theory Computations in Many-Fermion Systems},
  author = {He, Yuan-Yao and Shi, Hao and Zhang, Shiwei},
  journal = {Phys. Rev. Lett.},
  volume = {123},
  issue = {13},
  pages = {136402},
  numpages = {6},
  year = {2019},
  month = {Sep},
  publisher = {American Physical Society},
  doi = {10.1103/PhysRevLett.123.136402},
  url = {https://link.aps.org/doi/10.1103/PhysRevLett.123.136402}
}

@article{PhysRevB.78.041101,
  title = {Role of backflow correlations for the nonmagnetic phase of the $t\text{--}{t}^{\ensuremath{'}}$ Hubbard model},
  author = {Tocchio, Luca F. and Becca, Federico and Parola, Alberto and Sorella, Sandro},
  journal = {Phys. Rev. B},
  volume = {78},
  issue = {4},
  pages = {041101},
  numpages = {4},
  year = {2008},
  month = {Jul},
  publisher = {American Physical Society},
  doi = {10.1103/PhysRevB.78.041101},
  url = {https://link.aps.org/doi/10.1103/PhysRevB.78.041101}
}

@article{PhysRev.46.1002,
  title = {On the Interaction of Electrons in Metals},
  author = {Wigner, E.},
  journal = {Phys. Rev.},
  volume = {46},
  issue = {11},
  pages = {1002--1011},
  numpages = {0},
  year = {1934},
  month = {Dec},
  publisher = {American Physical Society},
  doi = {10.1103/PhysRev.46.1002},
  url = {https://link.aps.org/doi/10.1103/PhysRev.46.1002}
}

@article{PhysRevB.40.506,
  title = {Numerical study of the two-dimensional Hubbard model},
  author = {White, S. R. and Scalapino, D. J. and Sugar, R. L. and Loh, E. Y. and Gubernatis, J. E. and Scalettar, R. T.},
  journal = {Phys. Rev. B},
  volume = {40},
  issue = {1},
  pages = {506--516},
  numpages = {0},
  year = {1989},
  month = {Jul},
  publisher = {American Physical Society},
  doi = {10.1103/PhysRevB.40.506},
  url = {https://link.aps.org/doi/10.1103/PhysRevB.40.506}
}

@article{PhysRevLett.62.1407,
  title = {Phase diagram of the two-dimensional negative-U Hubbard model},
  author = {Scalettar, R. T. and Loh, E. Y. and Gubernatis, J. E. and Moreo, A. and White, S. R. and Scalapino, D. J. and Sugar, R. L. and Dagotto, E.},
  journal = {Phys. Rev. Lett.},
  volume = {62},
  issue = {12},
  pages = {1407--1410},
  numpages = {0},
  year = {1989},
  month = {Mar},
  publisher = {American Physical Society},
  doi = {10.1103/PhysRevLett.62.1407},
  url = {https://link.aps.org/doi/10.1103/PhysRevLett.62.1407}
}

@article{PhysRev.102.1189,
  title = {Energy Spectrum of the Excitations in Liquid Helium},
  author = {Feynman, R. P. and Cohen, Michael},
  journal = {Phys. Rev.},
  volume = {102},
  issue = {5},
  pages = {1189--1204},
  numpages = {0},
  year = {1956},
  month = {Jun},
  publisher = {American Physical Society},
  doi = {10.1103/PhysRev.102.1189},
  url = {https://link.aps.org/doi/10.1103/PhysRev.102.1189}
}

@inproceedings{he2016deep,
  title={Deep residual learning for image recognition},
  author={He, Kaiming and Zhang, Xiangyu and Ren, Shaoqing and Sun, Jian},
  booktitle={Proceedings of the IEEE conference on computer vision and pattern recognition},
  pages={770--778},
  year={2016}
}

@article{PhysRevLett.132.066002,
  title = {Emergent Superconductivity and Competing Charge Orders in Hole-Doped Square-Lattice $t\text{\ensuremath{-}}J$ Model},
  author = {Lu, Xin and Chen, Feng and Zhu, W. and Sheng, D. N. and Gong, Shou-Shu},
  journal = {Phys. Rev. Lett.},
  volume = {132},
  issue = {6},
  pages = {066002},
  numpages = {6},
  year = {2024},
  month = {Feb},
  publisher = {American Physical Society},
  doi = {10.1103/PhysRevLett.132.066002},
  url = {https://link.aps.org/doi/10.1103/PhysRevLett.132.066002}
}

@misc{jiang2025competitionchargedensitywavesuperconductingorders,
      title={Competition between charge-density-wave and superconducting orders on eight-leg square Hubbard cylinders}, 
      author={Hong-Chen Jiang and Thomas P. Devereaux and Steven A. Kivelson},
      year={2025},
      eprint={2511.18644},
      archivePrefix={arXiv},
      primaryClass={cond-mat.str-el},
      url={https://arxiv.org/abs/2511.18644}, 
}

@article{dosovitskiy2020image,
  title={An image is worth 16x16 words: Transformers for image recognition at scale},
  author={Dosovitskiy, Alexey and Beyer, Lucas and Kolesnikov, Alexander and Weissenborn, Dirk and Zhai, Xiaohua and Unterthiner, Thomas and Dehghani, Mostafa and Minderer, Matthias and Heigold, Georg and Gelly, Sylvain and others},
  journal={arXiv preprint arXiv:2010.11929},
  year={2020}
}

@PREAMBLE{
 "\providecommand{\noopsort}[1]{}" 
 # "\providecommand{\singleletter}[1]{#1}%" 
}

\end{document}